# Coincidence Rates for Photon Pairs in WDM Environment

Jungmi Oh, Cristian Antonelli, and Misha Brodsky

*Abstract*— We demonstrate the flexibility that a wavelength selective switch could offer to bandwidth provisioning of potential multi-user quantum key distribution networks based on entangled pair sources. We derive an analytical expression relating the coincidence detection rates of the photon pairs to the switch bandwidth characteristics. Experimentally measured coincidence rates verify the theory in three distinct network configurations.

## I. INTRODUCTION

QUANTUM key distribution (QKD) emerged as the most practical application of a new field of quantum information science [1]. Yet none of a few existing commercial products has been adopted by major telecom carriers. The current generation of QKD systems are point-to-point only and are based on transmission of weak coherent laser pulses between two distant stations. However, a system built on a different principle, that is on the distribution of entangled photons, could offer a possibility of a network scalable to many pairs of users [2-4]. A pair of entangled photon is a quantum object that has no analog in classical world. One property of an entangled pair is that the results of measurements performed at two distant locations on individual photons are correlated. For example, for photons appropriately entangled in their polarizations, two distant single photon detectors would register photons coincidentally if they are preceded by identically oriented polarizers. On the other hand, no simultaneous detection will be recorded if the polarizers are oriented orthogonally. Such a binary outcome of two measurements at two separate locations permits a protocol for building a shared cryptographic key [5].

Entangled photon pairs are usually created by conversion of pump photons in a nonlinear medium (either $\chi^{(2)}$ or $\chi^{(3)}$). Most popular in the case of fiber transmission, are the entanglement of polarizations [6,7] and time-bin entanglement [8]. Excellent sources for these two types of photon-entanglement have been made available in recent years [9,10]. For both entanglement types, the energy conservation law requires the frequency bands of photons within the same pair to be symmetric around some central frequency (related to the frequency of pump photons). Hence, for a *broadband* source of entangled photons positioned in a central office a number of users at paired peripheral nodes could be served by proper bandwidth allocation to each user. This bandwidth allocation could be done in a fixed fashion with a de-multiplexer in a central office spreading the spectrum to the peripheral nodes. Alternatively, a flexible bandwidth provisioning could be achieved, for example, by placing a wavelength selective switch (WSS) in a central office [11]. In this case the physical connection between the end users and the central office is done by fiber links and remains fixed, but the bandwidth allocation can be varied by controlling the WSS settings. This way, out of $N$ users each one can be connected to the others with only $N$ fibers.

The Eckert QKD protocol [5] requires two users to perform simultaneous measurements and record coincidental detections of photons. Correspondingly, one of the most relevant characteristics of a QKD system is the coincidence rate it is able to produce. Predicting coincident detection rates for pairs of entangled photons traversing WDM networks could be a daunting task for a QKD system designer or an operator. One obstacle is various transmission effects such as polarization mode dispersion [12,13]. But even leaving the transmission aside, different and, in general, frequency-dependent loss along the two photon paths presents sufficient challenge. In addition, WDM switching elements vary in spectral shape of the individual WDM channels, in inter-channel frequency spacing, and in loss of the individual channels [14].

In this paper we study coincidence detection rates of a pair of correlated photons, which are spectrally and physically separated by a WSS. While these photons are not entangled in the way described above, their spectral properties are similar to either polarization [6,7] or time-bin entangled [8] pairs. By describing the overall action of the WSS by its transfer function, we define probabilities of a photon transmission to a certain WSS port. This allows us to derive a simple analytical expression for coincidence probabilities for Poissonian statistics of photon pairs. We present several experimental results that verify our theory and demonstrate its versatility. In the first experiment, we setup our broadband source of correlated pairs to cover the telecom C-band. The WSS selects

Manuscript received April 21, 2010. (Write the date on which you submitted your paper for review.)

J. M. Oh was with the AT&T Labs., Middletown, NJ 07748 USA. She is now with the Bell Labs Seoul, Mapo-gu, Seoul,120-270, Korea
(e-mail: Jungmi.Oh@alcatel-lucent.com).

C. Antonelli is with the Department of Electrical and Information Engineering and CNISM, University of L'Aquila, L'Aquila 67040, Italy (e-mail: cristian.antonelli@univaq.it).

M. Brodsky is with the AT&T Labs., Middletown, NJ 07748 USA (corresponding author to provide phone: 732-420-9034; fax: 732-368-9433; e-mail: brodsky@research.att.com).

a spectrally symmetric pair of individual WDM channels and directs each of them to a chosen output port. Two fibers connect the output ports to single photon detectors (SPD). We measure the coincident counts for different selected pairs of WDM channels. Our second experiment is based on the free spectral range (FSR) periodicity of our WSS. We tune the spectrum of our photon pair source away from the C-band to simultaneously cover the S and L bands instead. Then we repeat the measurements over the same set of WSS configurations as before. In both cases our theory matches measured coincidence rates. Finally, we propose a method of in-situ characterization of our system based on our theory. We open a wide band of WDM channels to each WSS output port to dramatically increase photon counts. By simultaneously fitting measured coincidence together with individual detector counts with our expressions we obtain the average number of generated photon pairs and the efficiency of each detector. The extracted parameters agree well with the results of independent measurements.

## II. EXPERIMENTAL SETUP

Fig. 1 shows our experimental schematic. To approximate the practical networking environment, we assemble our setup entirely from fiber-based commercial off-the-shelf components. The setup consists of three main parts: a photon

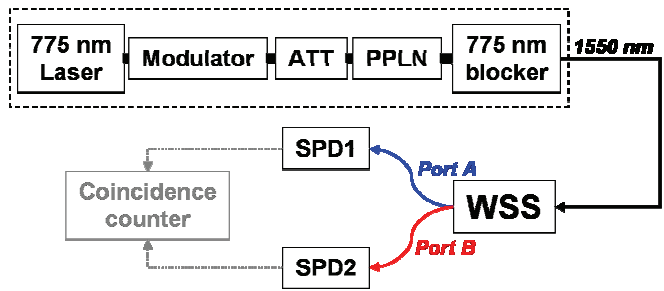

Fig. 1. Experimental setup.

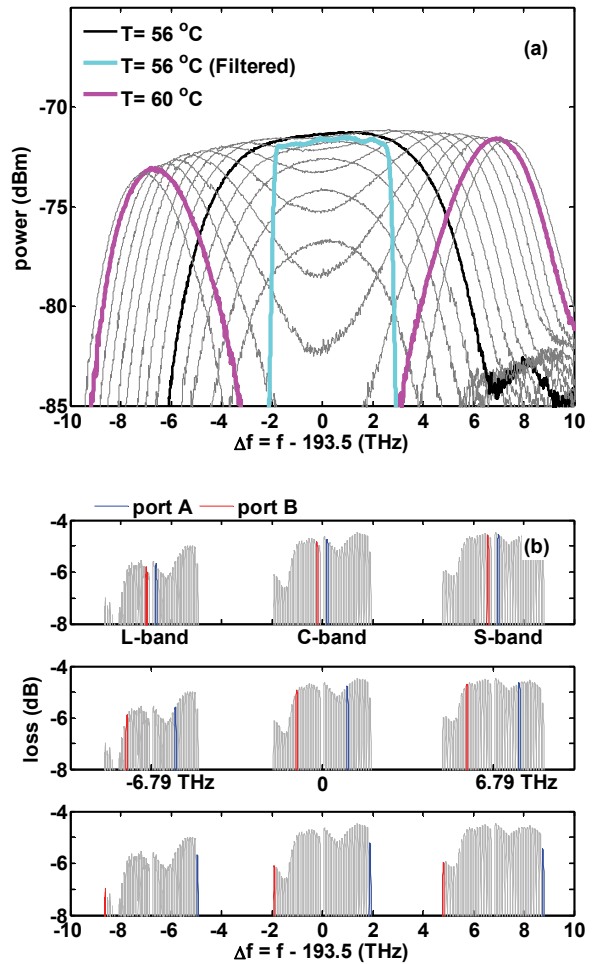

Fig. 2. (Color online) (a) A set of spontaneous down converted spectra each taken at different temperatures (grey). Spectrum taken at 56 ºC is marked by black line (unfiltered) or cyan (filtered). Spectrum taken at 60 ºC consist of two maxima and is marked by magenta color (resolution bandwidth 5 nm) (b) Frequency dependent loss of WSS. Three panels illustrate three different WSS configurations, in which one channel (blue) is directed to the output port A and the other one (red) to the port B. The channel number to the port A is 2 (top), 10 (middle), and 19 (bottom).

pair source, a WSS, and two SPDs with coincidence circuitry.

Telecom-band photon pairs are created by spontaneous down-conversion (SDC) in a periodically poled lithium niobate (PPLN) waveguide. We pump the waveguide with a cw semiconductor Fabry-Perot (FP) laser tuned by temperature to 774.66 nm to center the down-converted spectrum on 1549.32 nm (193.5 THz channel on 100 GHz-spaced ITU grid). Our FP operates in a quasi single mode regime, in which a dominant mode is about 24.5 dB stronger than its nearest neighbors. An external 10 Gb/s LiNbO$_3$ modulator sets a 1 MHz clock rate by carving a rectangular pulse from the cw lightwave. The active feedback bias control (not shown) maintains the extinction ratio of 23 dB. The maximal power conversion efficiency of our 20 mm-long PPLN waveguide is $1.6 \times 10^{-6}$ (when measured from fiber pigtail to fiber pigtail, including both input and output fiber coupling loss). This efficiency remains above $0.5 \times 10^{-6}$ for a wide temperature range of 55 ºC to 61 ºC. This corresponds to about 0.5 photon pairs over the entire spectrum per duration of the detector gate (the value is measured for zero value of attenuation ATT in Fig. 1).

The spectral properties of correlated photon pairs produced by downconversion have been extensively studied in the past [15,16]. Here for the sake of clarity we repeat some basic concepts. Our PPLN waveguide is kept at constant, yet arbitrarily selectable, temperature. The temperature controls resonant wavelength of the waveguide and thus the phase matching condition for the downconversion process. Correspondingly, the spectrum changes with temperature as illustrated in Fig. 2a. Here, we plot a set of spectra, each taken at a different temperature value, as the temperature was stepped from 54.0 ºC to 60.5 ºC. In our experiments we use two different waveguide temperatures. First, the degenerate spectrum at $T$=56 ºC provides the maximum power over the C-band (thick black line in Fig. 2a). For this case we also use an additional bandpass filter that reduced the spectrum to 4.6 THz to cover the C-band only (cyan line in Fig. 2a). Second, a temperature of $T$=60 ºC shifts the photon pairs spectrum into a combination of the S and L bands (magenta line in Fig. 2a). In each case the pump laser is blocked at the output of the PPLN by a WDM coupler with greater than 90 dB rejection of 775 nm light.

Wavelength selective switches are designed for adding and dropping channels to/from WDM data streams. To accomplish this, the WSS partitions its input spectrum into many WDM channel bands and directs them individually to chosen output ports. Our WSS (Metconnex, WSS5400) has 45 WDM channels on the 100 GHz-spaced ITU grid ranging from 191.6 THz to 196 THz, which could be directed to any of the 8 output ports [17]. We use the following convention for channel numbering: $N_{ch}(f) = 10 \times (f - 193.5)$, where f is in THz. Hence, the zero channel is at the center of the down-converted spectrum. A total of 38 channels (19 pairs) is available for coincidence measurements. Each channel is nearly flat top with 3dB band of 77 GHz. Figure 2b shows the transmission spectrum of the 38 channels (thin gray lines).

The two WSS output ports are connected by fibers to single photon counting benchtop receivers (Princeton Lightwave, PGA600). The receivers (SPD1 and SPD2 in Fig. 1) operate in a gated mode (gate duration ~0.5ns) and are synchronized with the modulator. They output an electrical TTL pulse for each detection event. These pulses are fed into the FPGA-based counter that measures both individual and the coincidence counts [18].

The dispersive element used in this WSS is a conventional arrayed waveguide grating (AWG), and hence there are multiple bands (marked as L, C, and S in Fig. 2b). Their center frequencies are separated by 6.79 THz - the FSR of the gratings [19]. Three panels illustrate three different WSS configurations. In each configuration we set only *one* channel ($N_{ch}^A$) to be directed to port A, and the other one ($N_{ch}^B = -N_{ch}^A$) to port B. However, because of the inherent periodicity, a total of *three* 77 GHz-wide channel replicas (one for each S, C, and L band) are directed to the corresponding fibers. These replicas are marked by blue (red) colors in the plot for output port A (B). Top, middle and bottom panels in Fig. 2b correspond to $N_{ch}^A = 2$, $N_{ch}^A = 10$, and $N_{ch}^A = 19$ respectively. Note that for the experimental condition $T$=56 ºC (filtered) only the C-band transmission window of WSS plays a role and the replicas belonging to S and L bands can be ignored. Conversely, for the condition T=60 ºC the C-band becomes immaterial as there is no downconverted light at those frequencies. In this case the photon frequencies are again symmetric about 193.5 THz, but each one falls into both of those two bands. That is, the photon in the port A (B) is comprised by two blue (red) channel replicas in the S and L bands

## III. COINCIDENCE PROBABILITY

In the absence of the dark counts the probability of coincident detection of two transmitted photons $P_{12}^0$ and the probability of detecting a transmitted photon in the *i*-th detector $P_i^0$ depend on the photon pair statistics. For our case the statistics are Poissonian and the following expressions for those probabilities can be derived:

$$P_{12}^0 = 1 - \exp(-\mu Q_1 T_1 \eta_1) - \exp(-\mu Q_2 T_2 \eta_2) \\ + \exp[-\mu(Q_1 T_1 \eta_1 + Q_2 T_2 \eta_2 - Q_{12} T_1 T_2 \eta_1 \eta_2)] \quad (1a)$$

$$P_i^0 = 1 - \exp(-\mu Q_i T_i \eta_i) \quad (1b)$$

Here $\mu$ is the total average number of photon pairs over the *entire* spectrum per detection gate, $\eta_i$ is the efficiency of the *i*-th detector and $T_i$ is the frequency-independent transmittance of the optical path between the pair creation and photon detection points excluding the transmittance through WSS itself. In our case $T_i$ mostly reflects the output fiber-coupling loss of the PPLN waveguide.

The quantities $Q_1$, $Q_2$ and $Q_{12}$ are introduced here to account for creation and routing of a photon in a certain *frequency band*. The probability density function for a photon pair to be generated at the offset frequencies $\pm\omega$ from one half of the pump frequency is the properly normalized down-converted spectrum $S(\omega)$:

$$\wp(\omega) = 2 S(\omega) \Big/ \int_{-\infty}^{+\infty} S(\omega) d\omega \quad (2)$$

We describe the overall action of the WSS by the transfer functions $H_{p1}(\omega)$ and $H_{p2}(\omega)$ relating the spectrum at the WSS input to the output spectra at WSS port 1 and port 2. These functions reflect a particular configuration of the WSS and change every time the WSS is reconfigured. Since conventional WSSs do not have a broadcasting capability, the two transfer functions do not overlap ($H_{p1}(\omega) H_{p2}(\omega) = 0$). Finally, the joint probability that each photon of the same pair is transmitted to the corresponding WSS ports is:

$$Q_{12} = \int_{-\infty}^{+\infty} \wp(\omega) |H_{p1}(\omega)|^2 |H_{p2}(-\omega)|^2 d\omega \quad (3a)$$

and the probability of a photon to appear at the *i*-th output port is given by:

$$Q_i = \int_{-\infty}^{+\infty} \wp(\omega) |H_{pi}(\omega)|^2 d\omega \quad (3b)$$

In the presence of dark counts coincident detections at the receiver could arise from either a photon pair generated in the PPLN, which is successfully transmitted through the fibers, or by dark counts or by the combination of the two. The total probability of a coincidence can be calculated as the complement to one of the total probability of several events. Using $P_{12}^0$ and $P_i^0$ of (1) and introducing $P_{dci}$ as the probability of a dark count in the *i*-th detector we obtain:

$$P_{12} = 1 - (1 - P_{dc1})(1 - P_{dc2})(1 - P_{12}^0) \\ - P_{dc1}(1 - P_{dc2})(1 - P_1^0) \quad (4a) \\ - P_{dc2}(1 - P_{dc1})(1 - P_2^0)$$

$$P_i = 1 - (1 - P_{dci})\exp(-\mu Q_i T_i \eta_i) \quad (4b)$$

As an aside, we would like to mention that the applicability of the presented methodology extends beyond the WSS context. In fact, it offers a description of any practical situation in which the loss of a quantum channel is frequency dependent. Interestingly, a closed form expression for $P_{12}^0$ exists even when the statistics of the pairs is many-fold thermal. It converges to the expression of (1a) as the number of modes goes to infinity.

## IV. EXPERIMENTAL RESULTS

First, we test the theory by measuring coincident counts between individual WDM channels over the C-band. We set the waveguide phase matching temperature to 56 °C, and use a filter to ensure the downconverted spectrum (cyan line in Fig. 2a) has no overlap with the S- and L-band of our WSS. A pair of symmetric channels $N_{ch}^A$ and $N_{ch}^B = -N_{ch}^A$, is directed to ports A and B, correspondingly, and the coincidence counts are measured for 30 seconds. Resulting coincidence probability is plotted as grey squares in Fig. 3a as a function of $N_{ch}^A$. The size of the counting error bars would be smaller than the symbol size The dependence is not flat mostly due to the wavelength dependent loss of the WSS. The independent measurements of the loss spectrum, permit evaluation of the integrals $Q_1$, $Q_2$ and $Q_{12}$. Using them together with independently obtained average number of pairs $\mu$ (see below), we calculate coincidence probability $P_{12}$ from (4a). The result is plotted in cyan in Fig. 3a. Interestingly, the theory fits the data well and captures nearly all features of a nontrivial frequency profile exhibited by the data.

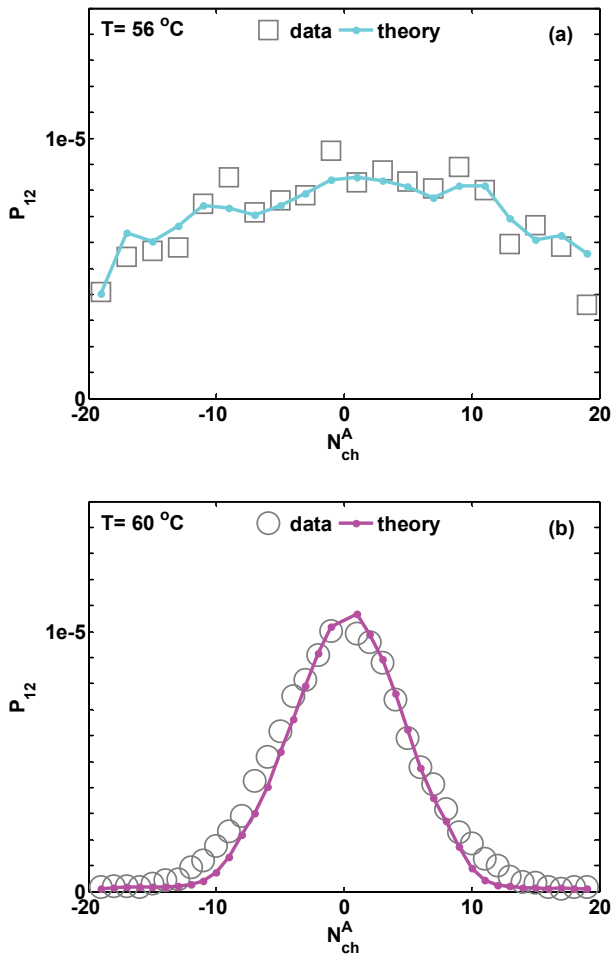

Fig. 3. (Color online) Coincidence probability $P_{12}$ as a function of channel number into port A. Symbols show the data, color lines the theory. (a) C-band, T=56 °C (b) S/L-band, T=60 °C.

The second test is slightly more subtle. It utilizes the FSR periodicity of one of our WSSs [19]. Now, for the same configuration of the WSS, that is for $N_{ch}^A$ and $N_{ch}^B = -N_{ch}^A$ being selected for the ports A and B, the actual frequency bands that are directed to each port consist not of one but of two well separated WDM peaks. The aggregate outputs are not completely symmetric in frequency and additionally they vary in loss. Indeed the *center frequencies* of three bands are separated by exactly the same amount $FSR = 6.79THz$, but the *channel separation* within each band is not the same. Our WSS has channel separation of 103.6 GHz, 99.9 GHz, and 96.5 GHz for the S, C, and L band, respectively. This behavior arises from the material properties and is common for all AWGs, on which the WSS is based on. So for large channel numbers ( $|N_{ch}^{A,B}| \geq 10$ ) the symmetry is lost and, therefore, the coincidence drops as can be seen in our data (grey circles in Fig. 3b). Again the theoretical prediction based on (4a) agrees well with the data. Interestingly, because the coincidence rate remains relatively high for at least a few central channels, the FSR periodicity property of AWGs together with the temperature tuning of the PPLN waveguide could be used to quickly move the quantum channel to and from the C-band without interrupting QKD service for a long period of time.

We have demonstrated that (4a) permits predicting the coincidence rates for a known average number of *created* photon pairs $\mu$. While $\mu$ might be known in lab experiments, in practical QKD systems an operator would be able to assess only the output power at the front panel of the entangled pair transmitter. This power differs from $\mu$ by a loss between the pair creation point and the transmitter output, which is unknown to the operator. However, this loss and, hence, $\mu$ can be determined if (4a-b) are considered together. Inset of Fig. 4a shows that 19 WDM channels are directed to each of the two WSS ports. We measure individual detector counts together with coincidence counts for several values of the 775-nm attenuator (ATT in Fig. 1). The measured probabilities are plotted by symbols in Fig. 4a for individual detectors and in Fig. 4b for coincidence. By performing simultaneous fits to the data based on the expressions in (4a-b) we obtain the detector efficiencies $\eta_i$ and the average number of created photon pairs over the entire spectrum during the detector gate $\mu$. The thin lines plot (4a-b) with fitted parameters, indicating the excellent quality of our fits. To diminish the counting errors and to speed up the measurement procedure we operate at relatively high power with multiple pairs being generated. As an illustration, the thin grey line in Fig. 4b plots would be coincidence probability in the absence of multiple pairs.

We perform several measurements in the C-band by using two WSSs from different manufacturers (Metconnex and Finisar) [17], [20]. The extracted detector efficiency is about 0.2. Each day the absolute value varied slightly presumably due to the room temperature variations. Yet on every day the extracted value was always within 0.01 of the results of independent reference measurements conducted on the same day. We also found the same consistency for the extracted average number of *created* photon pairs. By averaging results of different measurements we obtain $\mu = 0.5 \pm 0.06$ (over the cyan spectrum of Fig. 2a) for no attenuation in 775nm light path, that is ATT = 0. Interestingly, this value corresponds to the 3.2dB of fiber-optic coupling loss on 1550-nm side of our PPLN, which is extremely close to the manufacturer's estimates of the same loss of 1.5-3 dB.

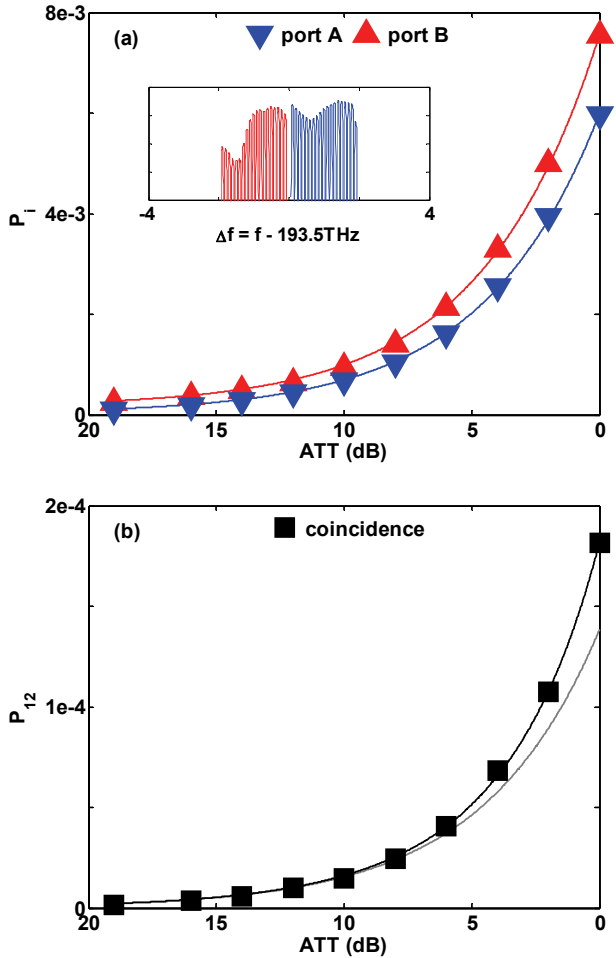

Fig. 4. (Color online) (a) Measured individual count probabilities $P_i$ (symbols) and relative fits (lines). Inset: WSS configuration. (b) Measured coincidence probability $P_{12}$ (symbols) and relative fit (black line). Grey line shows a would be coincidence rate in the absence of multiple pair generation.

## V. CONCLUSION

In this paper, we provide a method for calculating correlated photon coincidence rates in practical fiber optic networks, in which the loss is an arbitrary function of frequency. Experimentally measured coincidence rates verify our analytical expression in three different cases. We also suggest a possible way to use a WSS for provisioning a multi-user network based on an entangled photon source.